\documentclass{article}
\usepackage{spconfa4,amsmath,graphicx,amssymb}

\title{MIND THE MICROPHONE GAP: BENCHMARKING ARRAY UPSAMPLING\\ STRATEGIES FOR LATENT ACOUSTIC MAPPING}

\sixauthors
{{\em Philipp Schmidt}$^{1}$}
{{\em Huw Cheston}$^{1}$}
{{\em Juan Azcarreta}$^{2}$}
{{\em Adrian Stepien}$^{2}$}
{{\em Çağdaş Bilen}$^{2}$}
{{\em Iran R. Roman}$^{1}$}
{Queen Mary University of London}
{Meta Reality Labs Research}

\usepackage[hidelinks]{hyperref}
\usepackage{graphicx}
\usepackage{multirow}
\usepackage[table]{xcolor}
\usepackage{svg}
\usepackage{siunitx}
\usepackage{cleveref}
\usepackage{tabularx}
\usepackage{array}
\usepackage{ragged2e}
\usepackage{xcolor}
\usepackage{soul}

\usepackage{dblfloatfix}

\newcommand{\best}[1]{\textbf{#1}}
\newcommand{\second}[1]{\underline{#1}}

\newcolumntype{Y}{>{\RaggedRight\arraybackslash}X}

\begin{document}
\maketitle
\begin{abstract}
    Latent Acoustic Mapping (LAM) is a self-supervised learning method that generates high-resolution spherical acoustic maps from multichannel recordings without labelled data, matching supervised baselines on direction-of-arrival benchmarks.
    However, LAM degrades significantly with sparse 4-channel arrays, as the low-resolution cross-spectral matrix captures far less spatial information than the 32-channel inputs LAM was designed for.
    We benchmark a diverse set of upsampling architectures, spanning lightweight convolutional networks, iterative back-projection models, physics-informed networks, and generative adversarial approaches.
    We also study whether aligning these upsamplers with LAM by training them jointly or in different stages helps preserve the spatial structure that LAM depends on.
    Results show that the original full-resolution LAM is the strongest, that separately trained lightweight models are the most competitive learned approaches, and that representation alignment between the upsampler and LAM matters more than model complexity.
\end{abstract}
%


\section{Introduction}
\label{sec:Introduction}

The Latent Acoustic Mapping (LAM) model~\cite{lam} is self-supervised to generate high-resolution spherical acoustic maps (SAMs) directly from multichannel microphone recordings, without requiring labelled training data.
LAM encodes the cross-spectral matrix (CSM)~\cite{johnson1992} of a microphone array via a learnable back-projection~\cite{vanderveen2013} that maps covariance structure onto a Fibonacci tessellation~\cite{kushwaha2022}, producing an initial ``dirty'' image that is refined through four denoising convolutions~\cite{zhang2017,ilesanmi2021}.
The decoder then reconstructs the CSM from the denoised SAM using the array's steering matrix~\cite{krim1996}, and the model is trained end-to-end.
With only 16K parameters, LAM operates efficiently and in parallel across frequency bands, matching or exceeding supervised counterparts~\cite{simeoni2019,roman2024}. 

Three issues persist, however.
First, LAM is a high-resolution model (with optimal performance using a 32-channel array) and real-world SELD systems commonly use compact 4-channel tetrahedral arrays~\cite{shimada2023,adavanne2019} (due to cost and hardware constraints). The difference between the $4\times4$ and $32\times32$ CSM matrices is dramatic, with the former capturing drastically less spatial information.
A solution is to upsample the low-resolution CSM to the 32-channel space~\cite{lam, roman2024}.
Second, in light of the above it remains unclear which super-resolution architectures best preserve the spatial structure that LAM' depends on~\cite{simeoni2019,chardon2021}.
Third, even given a capable upsampler, the training strategy (whether to train the upsampler in isolation, in alignment with LAM, or fully end-to-end) critically affects downstream localisation performance~\cite{roman2024}, since joint training risks distorting the physical meaning of the CSM and domain overfitting.

This paper addresses all three points through a benchmark revealing how architecture and training strategy affect preservation of spatial information and downstream localisation.

\section{Approach}
\label{sec:Upsampling}

An upsampler maps a CSM $\mathbf{C}_\text{low} \in \mathbb{C}^{k \times M_\text{low} \times M_\text{low}}$\textemdash with frequency band index $k$ and low-resolution microphone count $M_\text{low}$\textemdash to a high-resolution target $\mathbf{C_\text{high}} \in \mathbb{C}^{k \times M_\text{high} \times M_\text{high}}$.
Because the CSM is complex-valued, all encoders are adapted to operate on its real and imaginary parts independently, thus approximating the Hermitian structure.
We therefore aim to determine which encoder best serves this role; the candidates are introduced below in chronological order based on their original publication, and are also summarised in
\Cref{table:model_comparison}.

\begin{table*}[]
    \centering
    \renewcommand{\arraystretch}{0.90}
    \setlength{\tabcolsep}{2pt}
    \resizebox{\textwidth}{!}{
        \begin{tabularx}{\textwidth}{@{} l Y r@{\;}r@{\;}l c c c @{}}
\hline
\textbf{Model} & \textbf{Description} & \multicolumn{3}{c}{\textbf{Parameters}} & \textbf{GFLOPs} & \textbf{Latency} & \textbf{Peak mem.} \\
\hline
\rowcolor{gray!10}
LAM~\cite{lam} & Reference full-resolution model (no upsampler). &  & 0.146 & M & 0.61 & \qty{0.48}{\ms} & \qty{1776}{\mega\byte} \\
\hline
Bicubic~\cite{bicubicInterpolation} & Interpolation with no learned parameters. & + & 0     & M & 0.61 & \qty{0.53}{\ms} & \qty{1808}{\mega\byte} \\
SRCNN~\cite{srcnn} & Three-layer CNN upsampling.
& + & 0.063 & M & 1.78 & \qty{0.60}{\ms} & \qty{1808}{\mega\byte} \\
DBPN~\cite{dbpn} & Iterative convolutional up- and down-sampling.
& + & 2.694 & M & 0.72 & \qty{0.62}{\ms} & \qty{1808}{\mega\byte} \\
IMDN~\cite{imdn} & Distillation network retaining key features for reconstruction. & + & 0.730 & M & 0.82 & \qty{0.51}{\ms} & \qty{1808}{\mega\byte} \\
SAFMN~\cite{safmn} & Feature-mixing network with spatially-adaptive modulation. & + & 0.772 & M & 0.83 & \qty{0.52}{\ms} & \qty{1808}{\mega\byte} \\
AINN~\cite{ainn} & Acoustics-informed MLP constrained by array geometry. & + & 0.007 & M & 0.74 & \qty{0.57}{\ms} & \qty{1876}{\mega\byte} \\
GAN~\cite{gan} & Adversarial generator to upsample HRTF measurement locations. & + & 4.287 & M & 4.95 & \qty{0.61}{\ms} & \qty{3443}{\mega\byte} \\
\hline
\end{tabularx}
    }
    \caption{The benchmarked upsampler models, including the parameter count that each adds to LAM (for the GAN, discriminator parameters are excluded as it is not used in inference). GFLOPs, latency and peak inference memory are reported for each including LAM. Peak inference memory is the median peak CUDA memory (see \Cref{sec:methodology}).
    }
    \label{table:model_comparison}
\end{table*}

\noindent\textbf{SRCNN:} The Super-Resolution Convolutional Neural Network (SRCNN) \cite{srcnn} is considered the first of its kind for image super-resolution.
It first upsamples the input to the target resolution via bicubic interpolation, then refines the result through three successive operations: overlapping patch extraction, non-linear feature mapping, and reconstruction~\cite{srcnn}.

\noindent \textbf{DBPN:} The Deep Back-Projection Network (DBPN) \cite{dbpn} consists of convolutional up- and down-projection blocks that alternate. Residual connections carry information across all preceding blocks of the same type.
The final high-resolution output is produced by concatenating the feature maps from every up-projection stage, allowing the network to exploit multi-depth representations of the input.
DBPN was used as the sole upsampler in the original LAM \cite{lam}.

\noindent \textbf{IMDN:} The Information Multi-Distillation Network (IMDN) \cite{imdn} reduces computation via information multi-distillation blocks, each of which progressively distils features and weighs their contribution via contrast-aware attention.
Residual connections link the blocks, and a sub-pixel convolution performs the final spatial upsampling, yielding a compact model with a smaller parameter count than DBPN. 

\noindent \textbf{SAFMN:} The Spatially-Adaptive Feature Modulation Network (SAFMN) \cite{safmn} is a lightweight model centred on global-local feature interaction.
At its core is a feature mixing module consisting of a spatially-adaptive modulation layer (depth-wise convolution followed by nearest-neighbour interpolation) and a convolutional channel mixer.

\noindent \textbf{AINN:} The Acoustics-Informed Neural Network (AINN) \cite{ainn} departs from the convolutional paradigm and is instead a multi-layer perceptron. Physics knowledge is injected  during training by penalising violations of the Helmholtz equation evaluated on the microphone array's geometry. Given the CSM's well-defined acoustic interpretation, the AINN stands out as a domain-specific encoder in our benchmark.

\noindent \textbf{GAN Generator:} The generator of a Generative Adversarial Network (GAN) can perform super-resolution for head-related transfer functions (HRTFs)~\cite{gan}.
The architecture consists of convolutional blocks that produce the final high-resolution output.
Adversarial training encourages the generation of more perceptually consistent representations, albeit at the cost of increased training complexity.

\noindent \textbf{Bicubic:} Used as a low-compute performance-floor reference point.
It estimates each output sample from a weighted combination of a $4 \times 4$ neighbourhood, yielding smoother transitions than bilinear or nearest-neighbour methods~
\cite{bicubicInterpolation}.

\section{Methodology}
\label{sec:methodology}

All models contain an upsampler, which serves as the primary experimental factor in our study, followed by LAM \cite{lam}.
In every experiment, we initialise LAM with the weights released with the original publication.
The upsamplers, in contrast, are adapted for our complex-valued CSM task and trained from scratch.
Below, we outline the data representations we use and the optimisation procedure for each model.\footnote{Explore the full source code: \href{doi.org/10.5281/zenodo.19735049}{https://doi.org/10.5281/zenodo.19735049}}

\subsection{Datasets}
\label{subsec:datasets}

To train, we use two types of high-resolution data: synthetic soundscapes that we generate using AudibleLight \cite{audiblelight}, and real-world recordings from EigenScape \cite{eigenscapeDataset}.
The training/validation set is partitioned with ratios of $0.89/0.11$ for AudibleLight and $0.86/0.14$ for EigenScape, which contain 570 minutes and 640 minutes of audio data, respectively.

Evaluation is done on three test sets.
First, we use a held-out AudibleLight set to test performance on unseen synthetic scenes.
Second, we test on LOCATA tasks 1\textendash 4 \cite{lollmann2018} (also high-resolution) and third, we test on STARSS23 \cite{shimada2023} (only available as low-resolution tetrahedral format).

In summary, our experimental use of data is therefore modeled after Roman et al.\ \cite{lam}, with two main differences. In addition to evaluating with LOCATA and STARSS23, we also evaluate using new AudibleLight test data. Also note that we completely exclude STARSS23 from model training.

\subsection{Training}
\label{subsec:training}

The upsampler is trained to reconstruct high-resolution 32-channel Eigenmike CSMs from a tetrahedral four-channel configuration, comprising Eigenmike microphones 6, 10, 22, and 26, consistent with previous literature \cite{lam,roman2024,adavanne2019,kushwaha2023sound}.
As the reconstruction is performed independently for each frequency band $k$, the upsampler learns the mapping $\mathbf{C}^{(k)}_\text{low}\in\mathbb{C}^{4\times 4}\rightarrow \mathbf{C}^{(k)}_\text{high}\in\mathbb{C}^{32\times 32}$.
The CSM produced by the upsampler serves directly as the input to LAM.

\begin{figure}[]
    \centering
    \includegraphics[width=\columnwidth]{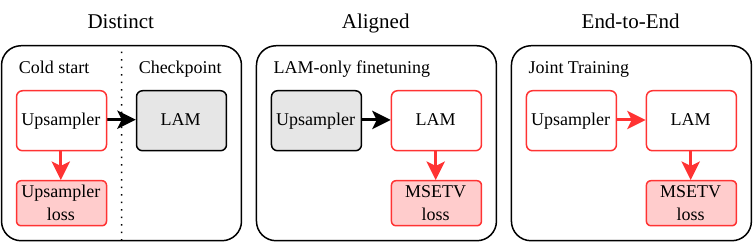}
    \caption{The three training strategies: rectangles are model components, red outlines indicate which are trainable while greyed-out ones have been trained previously, red arrows show the backpropagation loss.}
    \label{fig:training_experimental_design}
\end{figure}

We investigate three upsampler\textendash LAM training strategies, illustrated in \Cref{fig:training_experimental_design}.
In the \textit{distinct} setup, the upsampler and LAM are trained independently and never interact during optimisation.
Each upsampler is trained with its own reconstruction loss; 
The trained upsampler is then paired with LAM in the form released by its original authors.
In the \textit{aligned} setup, the upsampler trained in the \textit{distinct} stage is frozen, and only LAM is fine-tuned. The goal is to close the gap between upsampler and LAM.
The mean squared error and total variation loss (MSETV) is computed on the LAM output only.
In the \textit{end-to-end} setup, the upsampler and LAM are optimised jointly using the MSETV loss on the LAM output, with the upsampler remaining fully trainable throughout.

Across all experiments, the input audio, sampled at \qty{24}{\kilo\hertz}, is converted into a complex-valued CSM over nine linearly spaced frequency bands (\qtyrange{1.5}{4.5}{\kilo\hertz}). 
Optimisation uses AdamW ($\text{lr}=1.0\times10^{-4}$, $\text{weight decay}=1.0\times10^{-4}$, $\text{gradient clipping}=1.0$) with a batch size of 32, early stopping after 10 epochs without improvement on the validation loss, retaining only the best checkpoint.

\subsection{Evaluation}
\label{subsec:evaluation}

We evaluate in terms of localisation performance (including both localisation error and recall), computational efficiency, memory requirements, and CSM reconstruction fidelity.
For evaluation, all inference runs are carried out on an NVIDIA A100 40GB (training and inference are run on the same CUDA-enabled machine), a frame size of \qty{100}{\ms}, 4 data loader workers, and an inference batch size of 1.

The predicted spatial maps are converted into direction-of-arrival (DoA) estimates using adaptive k-means clustering, in which the number of active sources is determined from the structure of the predicted intensity peaks, consistent with Roman et al.~\cite{lam,roman2024}.
These DoA estimates are then evaluated using sound event localisation and detection (SELD) metrics.

Evaluation is performed in a class-agnostic setting, that is, source classes are ignored and all detections are treated as belonging to a single category.
K-means clusters with centroids within \ang{10} of each other are merged. Others stay separated.

For latency and peak CUDA memory allocation measurements, we perform 10 warm-up runs followed by 10 measurement runs (all on LOCATA), and we report the median over these forward passes.
To obtain the models' computational cost, we measure the floating point operations (FLOPs) on the full input tensor.

To assess CSM fidelity throughout the model, we compute the Correlation Matrix Distance (CMD)~\cite{herdinCorrelationMatrixDistance2005a} between the ground-truth 32-channel CSM and the CSMs produced at each stage: after the upsampler and after each of LAM's denoising steps. CMD is based on the Frobenius norm and ranges from 0 (perfect correlation) to 1 (full orthogonality).

\begin{table}[]
    \centering
    \renewcommand{\arraystretch}{1.00}
    \setlength{\tabcolsep}{3pt}
    \footnotesize
    \resizebox{\columnwidth}{!}{
        \newcommand{\tablerowspread}{\rule{0pt}{2.1ex}}
\begin{tabular}{l l|ccc|ccc}
\hline
\multirow{2}{*}{Model} & \multirow{2}{*}{Variant}
& \multicolumn{3}{c}{Loc. Error $\downarrow$}
& \multicolumn{3}{c}{Loc. Recall $\uparrow$}\\
\cline{3-8}
& 
& LOC \tablerowspread & S23 & AL
& LOC & S23 & AL \\
\hline
\multirow{2}{*}{Bicubic}
& \textit{Distinct}
& \ang{60.3} & \ang{59.3} & \ang{69.3}
& 0.97 & 0.84 & 0.90 \\
& \textit{Aligned}
& \ang{33.4} & \ang{58.8} & \ang{48.6}
& 0.72 & 0.63 & 0.91 \\
\hline

\multirow{3}{*}{SRCNN}
& \textit{Distinct}
& \best{15.6°} & \ang{59.4} & \ang{73.1}
& 0.97 & \best{0.86} & \best{1.00} \\
& \textit{Aligned}
& \second{\ang{19.0}} & \ang{59.0} & \ang{50.2}
& 0.73 & 0.66 & 0.83 \\
& \textit{E2E}
& \ang{20.9} & \ang{59.7} & \ang{48.8}
& 0.72 & 0.66 & 0.81 \\
\hline

\multirow{3}{*}{CDBPN}
& \textit{Distinct}
& \ang{35.7} & \ang{55.4} & \ang{68.0}
& 0.84 & 0.75 & 0.94 \\
& \textit{Aligned}
& \ang{27.1} & \second{\ang{55.1}} & \ang{47.2}
& 0.73 & 0.70 & 0.92 \\
& \textit{E2E}
& \ang{19.4} & \ang{58.9} & \second{\ang{45.3}}
& 0.73 & 0.66 & 0.80 \\
\hline

\multirow{3}{*}{IMDN}
& \textit{Distinct}
& \ang{27.3} & \best{54.5°} & \ang{61.2}
& 0.91 & 0.81 & \second{0.99} \\
& \textit{Aligned}
& \ang{27.3} & \ang{57.2} & \ang{46.9}
& 0.72 & 0.66 & 0.84 \\
& \textit{E2E}
& \ang{22.7} & \ang{57.0} & \ang{48.5}
& 0.74 & 0.66 & 0.82 \\
\hline

\multirow{3}{*}{SAFMN}
& \textit{Distinct}
& \ang{20.8} & \second{\ang{55.1}} & \ang{58.9}
& 0.90 & 0.74 & 0.98 \\
& \textit{Aligned}
& \ang{24.6} & \ang{58.1} & \ang{45.8}
& 0.72 & 0.66 & 0.84 \\
& \textit{E2E}
& \ang{23.2} & \ang{57.0} & \best{44.5°}
& 0.72 & 0.66 & 0.80 \\
\hline

\multirow{3}{*}{AINN}
& \textit{Distinct}
& \ang{29.9} & \ang{62.1} & \ang{67.8}
& \second{0.99} & 0.84 & \best{1.00} \\
& \textit{Aligned}
& \ang{19.5} & \ang{58.8} & \ang{54.5}
& 0.74 & 0.67 & 0.83 \\
& \textit{E2E}
& \ang{21.3} & \ang{60.2} & \ang{52.3}
& 0.72 & 0.66 & 0.82 \\
\hline

\multirow{3}{*}{GAN}
& \textit{Distinct}
& \ang{28.2} & \ang{63.9} & \ang{81.3}
& \best{1.00} & 0.84 & 0.98 \\
& \textit{Aligned}
& \ang{31.9} & \ang{55.3} & \ang{81.2}
& 0.73 & 0.67 & 0.85 \\
& \textit{E2E}
& \ang{27.0} & \ang{62.8} & \ang{86.7}
& \second{0.99} & \second{0.85} & 0.93 \\
\hline

\rowcolor{gray!10}
LAM (ref.) \tablerowspread & --
& \ang{14.5} & -- & \ang{25.3}
& 0.75 & -- & 0.82 \\
\hline

\end{tabular}
    }
    \caption{Localisation error and recall on LOCATA (LOC), STARSS23 (S23), and AudibleLight (AL) data for each upsampler + LAM combination and training strategy. In each column, bold and underlined values denote the best and second-best, respectively. The greyed-out row shows full-resolution LAM performance on LOC and AL, for reference (S23 is only available in low resolution).}
    \label{table:results}
\end{table}

\section{Results}
\label{sec:results}

\Cref{table:results} shows the quantitative results.
Two findings stand out.

\begin{figure*}[]
    \centering\includegraphics[width=0.9\textwidth]{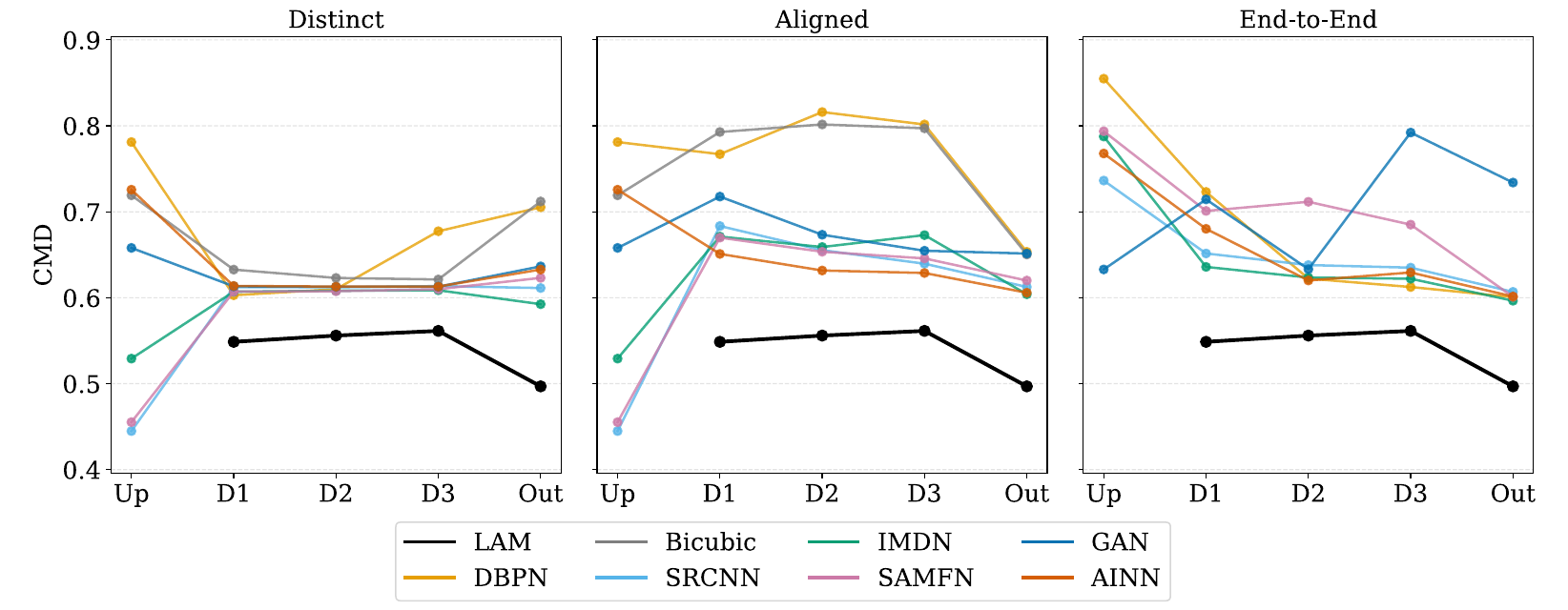}
    \caption{Correlation matrix distance (CMD; lower is better) between the different CSMs that exist at different stages of the LAM model and the ground truth CSM. \textit{Up} is the upsampler's CSM (the input to the full-resolution LAM variant, black line, is the ground truth CSM). \textit{D1}\textendash\textit{D3} are the CSMs decoded from the ``dirty'' latent acoustic images inside LAM, while \textit{Out} is LAM's output CSM.
    Data computed on the LOCATA dataset.}
    \label{fig:locata_cmd}
\end{figure*}

Across all upsamplers and datasets, \textit{distinct} training consistently yields the highest recall within each type. However, this benefit does not extend to localisation error. On LOCATA, only the SRCNN upsampler allows LAM to approach its full-resolution performance baseline (\ang{15.6} vs.\ \ang{14.5}); for all other upsamplers, \textit{distinct} training yields higher errors than their \textit{aligned} or \textit{end-to-end} counterparts. On AudibleLight, \textit{distinct} errors are consistently the highest within each model class. This is explained by the fact that in \textit{distinct} training LAM is kept frozen from its original version, and AudibleLight data is completely new and out of domain. 

Both \textit{Aligned} LAM and \textit{end-to-end} training of upsampler + LAM generally reduce localisation error, most clearly on AudibleLight where SAFMN+LAM (\textit{end-to-end}) and DBPN+LAM (\textit{end-to-end}) achieve the two lowest errors overall (44.5° and 45.3°). On LOCATA, recall is impacted to around 0.72\textendash0.74, regardless of model capacity. This recall-localisation trade-off is consistent and suggests that closer adaptation to LAM narrows the output distribution, but limits detection sensitivity.
Computational overhead does not predict performance; despite being the most expensive model, the GAN does not perform best, and there is no consistent trend between GFLOPs or latency and localisation error.

\Cref{fig:locata_cmd} shows that CSM fidelity explains downstream performance. For \textit{distinct} and \textit{aligned} training, CSM quality is directly related to localisation error.
For instance, full-resolution LAM (black line) yields CMD values below 0.6. In contrast, all upsampler + LAM combinations appear to hit a floor around 0.6. Importantly, upsamplers like SRCNN, SAFMN, and IMDN, seem capable of producing CSMs even better than LAM's output.
This indicates that bridging the latent acoustic image and the CSM is difficult or that LAM is not the optimal model to carry this out.
In the \textit{end-to-end} case, the upsampler's output CSM drifts further from the true CSM, indicating that the upsampler learns a LAM-specific representation rather than a faithful reconstruction. This makes sense as these models were supervised using only the LAM loss.

\section{Conclusions}
\label{sec:conclusions}

We presented the first  benchmark of array upsampling strategies for Latent Acoustic Mapping, evaluating six architectures across different training strategies and datasets. Three findings stand out.
Model complexity does not predict localisation quality: despite 60× fewer parameters, SRCNN matches or outperforms the GAN, and no consistent relationship exists between GFLOPs and localisation error. Training strategy matters more than architecture: distinct training maximises recall but inflates localisation error, while joint and aligned training reverse this trade-off, a pattern consistent enough across models and datasets to suggest it is a structural property of the LAM pipeline. Finally, our staged CMD analysis explains why end-to-end training underperforms: the upsampler drifts from physically faithful CSM reconstruction toward a LAM-specific latent, a tension invisible without the intermediate diagnostic we introduce here.

Still, the metrics used do not support detailed analysis of the recall saturation on the STARSS23 dataset, which may indicate that false positives are not adequately captured. Moreover, our localisation measurements provide no measures of statistical uncertainty, such as variance or confidence intervals, limiting the statistical robustness of the results.

Our findings reframe the upsampling problem: priority should be matching the upsampler's output distribution to LAM's expected input, not maximising standalone reconstruction capacity. Future work should explore alignment objectives that explicitly enforce CSM structure, and whether other array geometries can reduce the upsampling burden.

\section{Acknowledgements}

P. Schmidt (QMUL) thanks Meta Platforms Inc. for financial support to travel and present this work at IWAENC 2026. 

\vspace{-8pt}

\bibliographystyle{IEEEbib}
\bibliography{refs}

\end{document}